\documentstyle[12pt]{article}

\newcommand{\beq}{\begin{eqnarray}}
\newcommand{\eeq}{\end{eqnarray}}

\begin{document}

\rightline{UTHEP-307}

\rightline{June, 1995}

\vskip 2cm

\vspace{5pt}
 \centerline{\Large{\bf  $\phi$-meson in Nuclear Matter}}

\vspace{1.0cm}
\centerline{H. Kuwabara and T. Hatsuda}

\vspace{0.5cm}

\noindent
 \centerline{\em Institute of Physics, University
 of Tsukuba, Tsukuba, Ibaraki 305, Japan}

\vspace{2cm}

\centerline{Abstract}
  $\phi$-meson mass in nuclear matter ($m_{\phi}^*$)
 is investigated using  an effective Lagrangian of $\phi$ interacting
 with octet baryons.
  $m_{\phi}^*$  decreases by a few \% in nuclear matter due to the
  current conservation and effective nucleon/hyperon masses.
  Its implication to the p-A and A-A collisions are briefly discussed.

\vspace{2cm}
\noindent
Submitted to Prog. Theor. Phys.

\newpage

\setcounter{equation}{0}
\renewcommand{\theequation}{\arabic{equation}}

In recent years, the effecitve meson masses in
 hadronic matter at finite density
 ($\rho$) and  temperature ($T$)
 acquire wide interests both theoretically and experimentally
 (see the recent reviews, \cite{yukawa}.)
   In particular, the  $\phi$-meson,
 which is a  $\bar{s}s$  resonance in 	$J^P = 1^-$ channel
 with  a narrow width
  ($m_{\phi} =1019.4$  MeV
 and $\Gamma_{\phi} = 4.4$ MeV), is a unique probe for
  partial restoration of chiral symmetry
   in hot/dense hadronic matter \cite{finiteT,LH,theory}.
  Detection of $\phi$ through the dacays
 $\phi \rightarrow K \bar{K}, e^+e^-, \mu^+ \mu^-$
  in nucleous-nulceous (A-A) and
 proton-nucleous (p-A)  collisions could give experimental
 information on the spectral change of $\phi$ in matter:
 preliminary data on $\phi \rightarrow K^+K^-$
  in $A-A$ colisions at AGS-BNL has been  recently reported \cite{AGS},
  and an  experiment using p-A reactions is planned at KEK \cite{Enyo}.

In this letter, we will report our recent study
 on the $\phi$-meson mass in nuclear matter at zero $T$.
 Our starting point is an effective hadronic model
 composed of $\phi$-meson, nucleon and
 hyperons. This is a generalization of the recent works
 by Shiomi and Hatsuda \cite{SH} and others \cite{others,french,PW}
  who studied the
 effect of nucleon-loops to the proporties of
  rho and omega mesons in nuclear matter.

Let's start with the vector coupling of $\phi$ with
 octet baryons
 ($B \equiv N, \Lambda^0, \Sigma^{\pm}, \Sigma^{0}$);
\beq
L_{int} = \sum_B g_{\phi B} \  \bar{B} \gamma_{\mu}B\  \phi^{\mu} ,
\eeq
where $g_{\phi B}$ being the $\phi$-baryon coupling constant listed in
 Table 1.

\vspace{0.8cm}

\begin{center}

\begin{tabular}{|c|c|c|c|} \hline
Baryons     & $g_{\sigma B}$  & $g_{\omega B}$  &
 $g_{\phi B}$   \\ \hline \hline
 N          &8.7      &10.6    &4.2$^*$     \\ \hline
 $\Lambda$  &5.2      &6.9     &2.3         \\ \hline
 $\Sigma$   &5.2      &6.9     &2.3         \\ \hline
\end{tabular}\\

\vspace{0.5cm}

\end{center}

\noindent
Table1:
 $g_{\sigma B}$, $g_{\omega B}$ and $g_{\phi B}$ denote
 $\sigma$-$B$ scalar coupling,
  $\omega$-$B$ vector coupling,
  and
  $\phi$-$B$ vector coupling, respectively.
  $g_{\sigma B}$ and  $g_{\omega B}$
 are taken from \cite{gled}.  The number with $^*$
  should be considered as an  upper bound.

\vspace{0.8cm}

 Some remarks are in order here:
  (i) $\phi - \Lambda$ and $\phi - \Sigma$ couplings
 do not break the OZI rule, since the quark lines at the vertices are
 connected. On the other hand, the $\phi - N$ coupling is OZI violating.
 (ii)  $\Xi$ is neglected, since its effect to the
 $\phi$ self-energy is doubly
 suppressed by the mass of $\Xi$ and by the OZI violating
 nature of $\phi-\Xi$ coupling.
 (iii) If one relies on the quark counting rule \cite{QM},
  the $\phi$-hyperon couplings  are related to the
 $\omega$-hyperon  couplings as
 $g_{\phi \Lambda (\phi \Sigma)} = g_{\omega \Lambda (\phi \Sigma)}/3$ with
 $g_{\omega \Lambda (\phi \Sigma)}$ being determined by the
 fit of the hypernuclear levels \cite{gled}. This is assumed in Table 1.
  (iv) $\phi$-nucleon coupling, which is OZI violating,
  is not known experimentally.
  However,  a study of the electromagnetic form-factors of the nucleon
 yields an upper bound of its strength \cite{jaffe}.
  Using the notation of ref.\cite{jaffe},
 $g_{\phi N}/g_{\omega N}= (\sin \epsilon + \cos \epsilon \tan \eta_1)
 /(\cos \epsilon - \sin \epsilon \tan \eta_1) \simeq \epsilon +
 \tan \eta_1 <  0.4 $.

We will consider only the $N=Z$ non-strange nuclear matter in this letter.
  In this case,  effects of the
 hyperons to the $\phi$-meson self-energy arise only through
  hyperon$-$anti-hyperon loops. Effective masses of hyperons
 $M_{\Lambda,\Sigma}^*$ in nuclear matter
 give density dependence of the self-energy.
    Nucleon contribution to the self energy
  has both $N-\bar{N}$ loop (polarization of the Dirac sea)
 and the scattering with nucleons in nuclear
 matter (polarization of the Fermi sea) \cite{others,PW,SH}.
   The one-loop self energy from hyperon and nucleon contributions
 reads
\beq
\label{polarization}
\Pi^{\mu \nu}_B (\omega, {\bf q};\rho)
 =- ig_{\phi B}^2 \int {d^4k  \over (2\pi)^4}\
{\rm Tr}[\gamma^\mu G(k+q) \gamma^\nu G(k)]  \ \ ,
\eeq
where the four  momentum of $\phi$ is
 $q^{\mu} = (\omega, {\bf q})$, ``Tr'' is for the Dirac indices,
 and
 $G(k)$ denotes  baryon propagator in nuclear matter
 which depends on the effective mass of the nucleon and hyperons
 $M_B^*$ ($B= N,\Lambda, \Sigma)$.  (See  \cite{SH} for
 the explicit form of $G(k)$.)

  Although one can calculate the density dependence
 of $M^*_B$ within the framework of quantum hadrodynamics (QHD)
 \cite{QHD}, we adopt the following forms to study
correlations between $M^*_B$ and $m_{\phi}^*$ in a qualitative manner:
\beq
\label{baryon}
 g_{\sigma \Lambda (\sigma\Sigma)}/g_{\sigma N}
 & = & ( M_{\Lambda (\Sigma)} -M_{\Lambda (\Sigma)}^*)
/ (M_N - M_N^*) \\
\label{nucleon}
M_N^*/M_N & \simeq & 1 - 0.15 \rho/\rho_0 ,
\eeq
where $\rho_0$ is the normal nuclear-matter density,
 and $g_{\sigma \Lambda (\sigma \Sigma)}$ and $g_{\sigma N}$ are
 given  in Table 1. Eq.(\ref{baryon}) is an
 universal relation in the relativistic mean field theory
 \cite{gled}
 and Eq.(\ref{nucleon}) is a standard parametrization
 for the the effective nucleon mass at $\rho < 2 \rho_0$ \cite{french}.
 	In Fig. 1, effective masses of $N$, $\Lambda$ and  $\Sigma$
 parametrized by eqs. (\ref{baryon},\ref{nucleon})
 are shown as a function of baryon density.

 The effective $\phi$-meson mass $m^*_{\phi}$
 at rest ($\omega \neq 0$, ${\bf q}=0$)
 is obtained as a  solution of the dispersion relation
\beq
\label{dispersion}
 \omega^2 - m_{\phi}^2 + \sum_B
\tilde{\Pi}_B (\omega, {\bf 0};\rho)   =0,
\eeq
where
$\tilde{\Pi}_B (\omega, {\bf 0};\rho) \equiv -
 \tilde{\Pi}_B^{\mu\mu}(\omega, {\bf 0};\rho)/ 3\omega^2 $, and
 $m_{\phi}$ is the $\phi$-meson mass in the vacuum.
 $\tilde{\Pi}_B^{\mu \mu}$ denotes the
  density dependent part of ${\Pi}_B^{\mu \mu}$:
  the density independent logarithemic divergence
  is subtracted out in the dimensional regularization scheme
  following the procedure
  given in \cite{french,SH}, namely
 $ \tilde{\Pi}_B^{\mu \mu}(\omega, {\bf 0};\rho)
 \equiv \Pi_B^{\mu \mu}(\omega, {\bf 0};\rho) -
 \Pi_B^{\mu \mu}(\omega, {\bf 0};0)$.
  On should note here that
 similar ``renormalized'' self-energy in the relativistic $\sigma-\omega$
 model has been used to study the  electron-nucleous scattering, and
 it has been shown that the renormalized Dirac-sea polarization
  gives rather good agreement with experiments on
 the quenching of the Coulomb sum values \cite{KS88}.

The solid line in Fig.2 show
 the ratio $m_{\phi}^*/m_{\phi}$ calculated in the above
 subtraction procedure with hyperon-loops only.
 The hyperon contribution is less ambiguous comapred to the
 nucleon contribution, since the absolute value of
  $g_{\phi N}$ in the latter case is not known.
 $m_{\phi}^*$ decreases by $2-3 \% $ in the range
 $\rho_0 < \rho < 2 \rho_0$.
 Note that
  the OZI rule is preserved for $\phi$-hyperon vertices, while it is
 violated in the
 self-energy  $ \tilde{\Pi}_B^{\mu \mu}(\omega, {\bf 0};\rho)$.
  This is because the self-energy
  respresents interaction
 of $\phi$ ($s\bar{s}$ pair)  with {\em non-strange} nuclear matter.
 Similar phenomena are known in two-step decay processes
 such as $\phi \rightarrow K\bar{K} \rightarrow \rho \pi$,
 $f' \rightarrow K\bar{K} \rightarrow \pi \pi$, and
$J/\psi \rightarrow D\bar{D} \rightarrow  \rho \pi$, where
 each vertex preserves the rule while
  the whole amplitude violates the OZI rule \cite{lipkin}.

 The solid line in Fig.3 shows
 the ratio $m_{\phi}^*/m_{\phi}$ calculated in the above
 subtraction procedure with nucleon-loops only.
  We have used $g_{\phi N}/g_{\omega N}
 = 0.32$ in Fig.3 which is close to the upper bound
 given in Table 1: thus the
 resultant decrease of $m_{\phi}^*$ in Fig.3 should be considered as
  an upper
 limit originating from the nucleon-loop. Note here that
 the nucleon contribution contains the polarization of Dirac sea
 and fermi sea.
  The fomer (latter) tends to decreases
 (increases) the effective mass.

 The nagitive mass shift in Fig.2 and Fig.3  is
 a direct consequence of the current conservation
 ($\partial_{\mu}(\bar{B}\gamma^{\mu}B)=0)$ and
 $M^*_B/M_B < 1$,
  which was first discussed in \cite{SH,PW} for rho and omega
 mesons.  For the $\phi$-meson,
  the current conservation implies that
 the propagator of $\phi$ (without a small Fermi-sea polarization)
  has a form $D(q) \simeq 1/(Z^{-1}q^2 - m_{\phi}^2)$
 with $Z$ being a {\em finite} wave-function renormalization in medium.
  $M_B^*/M_B < 1$ implies that $\phi$ is more dressed by
 baryonic clouds in medium, which leads to $Z < 1$.
  Thus, one arrives
 at the  conclusion  $m_{\phi}^*/m_{\phi} \equiv Z < 1$.
  This mechanism  is quite general and does not depend on the
 details of the interaction and on the virtual particles running
 in the loop;
 for example,  a similar decrease of $m_{\phi}^*$ should be seen even when
 one replaces the baryonic loops  by the
 constituent-quark loop.

 To see the effect of the ultraviolet cutoff on the {\em finite
 part} of the loop
 integral in (\ref{polarization}),
  let us define
  $ \tilde{\Pi}_B^{\mu \mu}(\omega, {\bf 0};\rho,\Lambda_{cut})
 \equiv \Pi_B^{\mu \mu}(\omega, {\bf 0};\rho,\Lambda_{cut}) -
 \Pi_B^{\mu \mu}(\omega, {\bf 0};0,\Lambda_{cut})$ and use this
 in (\ref{dispersion}). We take  covariant cutoff for
 $\Lambda_{cut}$ for simplicity.
  When $\Lambda_{cut} \rightarrow \infty$,
  $ \tilde{\Pi}_B^{\mu \mu}(\omega, {\bf 0};\rho,\Lambda_{cut})$
 reduces to $\tilde{\Pi}_B^{\mu \mu}(\omega, {\bf 0};\rho)$.
 The dashed lines in Fig.2 and Fig.3
 are the results of such calculations for three cases,
 $\Lambda_{cut}$ = 1, 2, 10  GeV.
  Although the cutoff dependence is not negligible,
 the qualitative picture we draw in the above is not affected.
 Also, ``renormalized'' curve (solid line) is more favorable
 from the phenomenological point of view, since it can explain
 the quenching of the Coulomb sum values as we have mentioned before.

We have considered only the nucleon and hyperon loops in the $\phi$
 self-energy.  Another possible contribution is the kaon-loop in medium,
 which was studied by Ko et al. \cite{theory}.
   They found that the kaon-loop also has a tendency to decrease
  $m_{\phi}^*$ at low densities provided that the
 effective kaon mass $\bar{m}_K^* = ( m^*_{K^{-}}+ m^*_{K^{+}})/2$
  decreases in medium.
  However, it is still controvertial whether $\bar{m}_K^*$ really
 decreases
 in nuclear matter or not (see e.g. \cite{Yabu}).
  In QCD sum rules (QSR), $m_{\phi}^*$ is shown to decrease as a
  result of the
 partial restoration of chiral symmetry in nuclear matter, in
  particular by
 the medium modification of the strangeness condensate
  $\langle \bar{s}s \rangle$ \cite{LH}.
  Unfortunatelly, it is hard to make a solid connection of
 this result with that in this letter, since the kinamatical region
 to extract $m_{\phi}^*$ in two approaches are quite different (
deep Euclidian region in QSR versus on-shell region in the approach here).

 Recently,  Enyo et al. have proposes an
experiment to create $\phi$-meson in heavy nuclei using the proton-nucleus
reaction and to detect lepton pairs and kaon pairs from $\phi$
 decaying in the nucleous \cite{Enyo}. A possible signal
 in this experiment is a double $\phi$-meson peak in the $e^+e^-$
 spectrum and also a large change of the branching ratio
 $\Gamma(\phi \rightarrow e^+e^-)/\Gamma(\phi \rightarrow K^+K^-)$.
   Also, E859 at BNL-AGS has recently reported
  a possible spectral change of the $\phi$-peak in
$K^+K^-$spectrum in heavy ion collisions \cite{AGS}.
 In such experiments, a shoulder structure of the $\phi$-peak
  should be a possible signal of the mass shift of $\phi$.

\vspace{3cm}

The authors thank H. Shiomi for useful discussions and helps.
 We also thank Y. Akiba for suggesting us to study
 the cutoff dependence, and H. Hamagaki, Y. Miake and K. Yagi
 for discussions on E859 data at AGS.

\vspace{2cm}

\centerline{\bf{Figure captions}}

\vspace{1cm}

\noindent
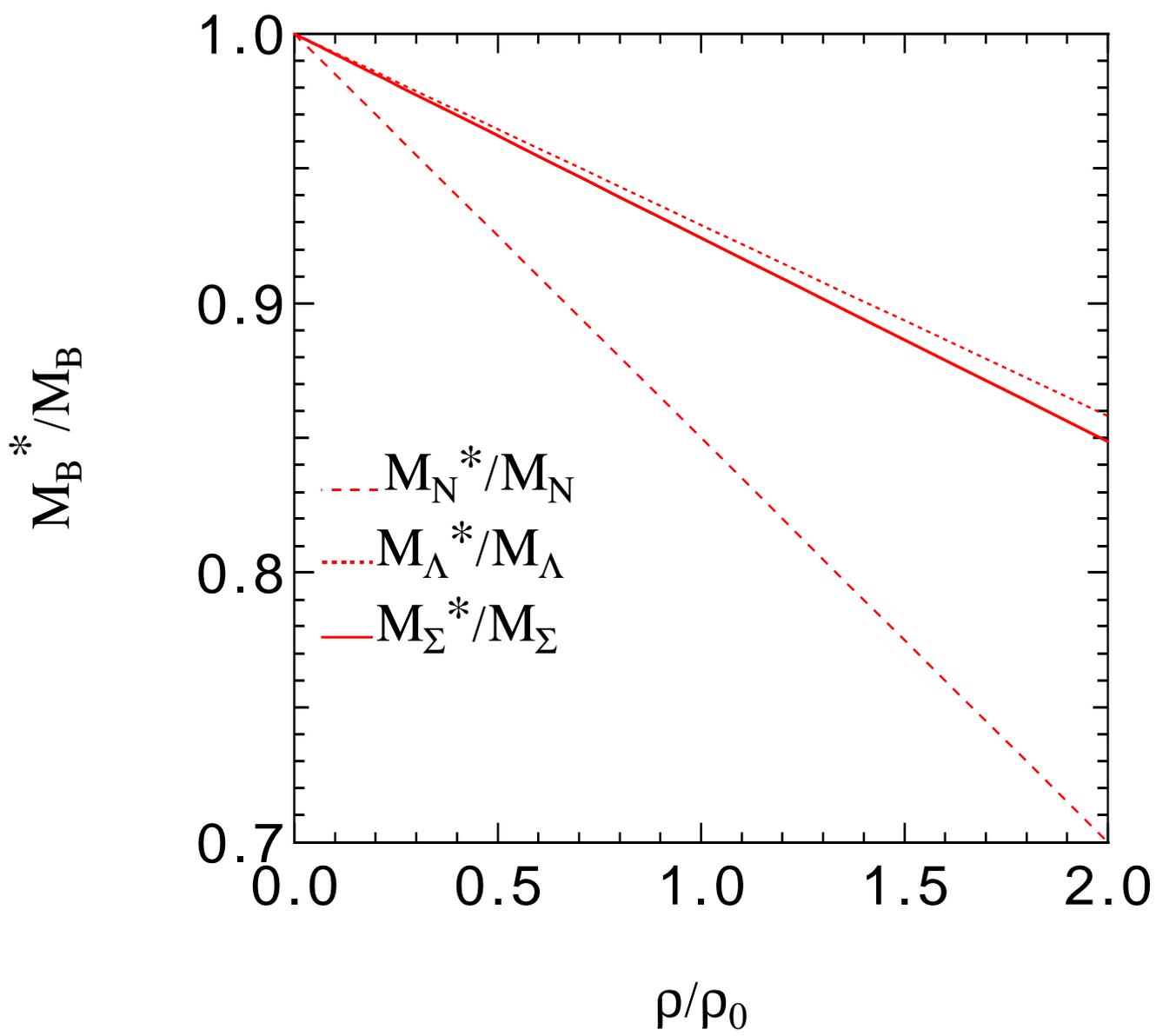
Fig.1: Ratio of baryon mass in matter $M_{B}^*$ and that in vacuum
 $M_{B}$ ($B=N, \Lambda, \Sigma$) as a function of  $\rho/\rho_0$.

\vspace{0.8cm}

\noindent
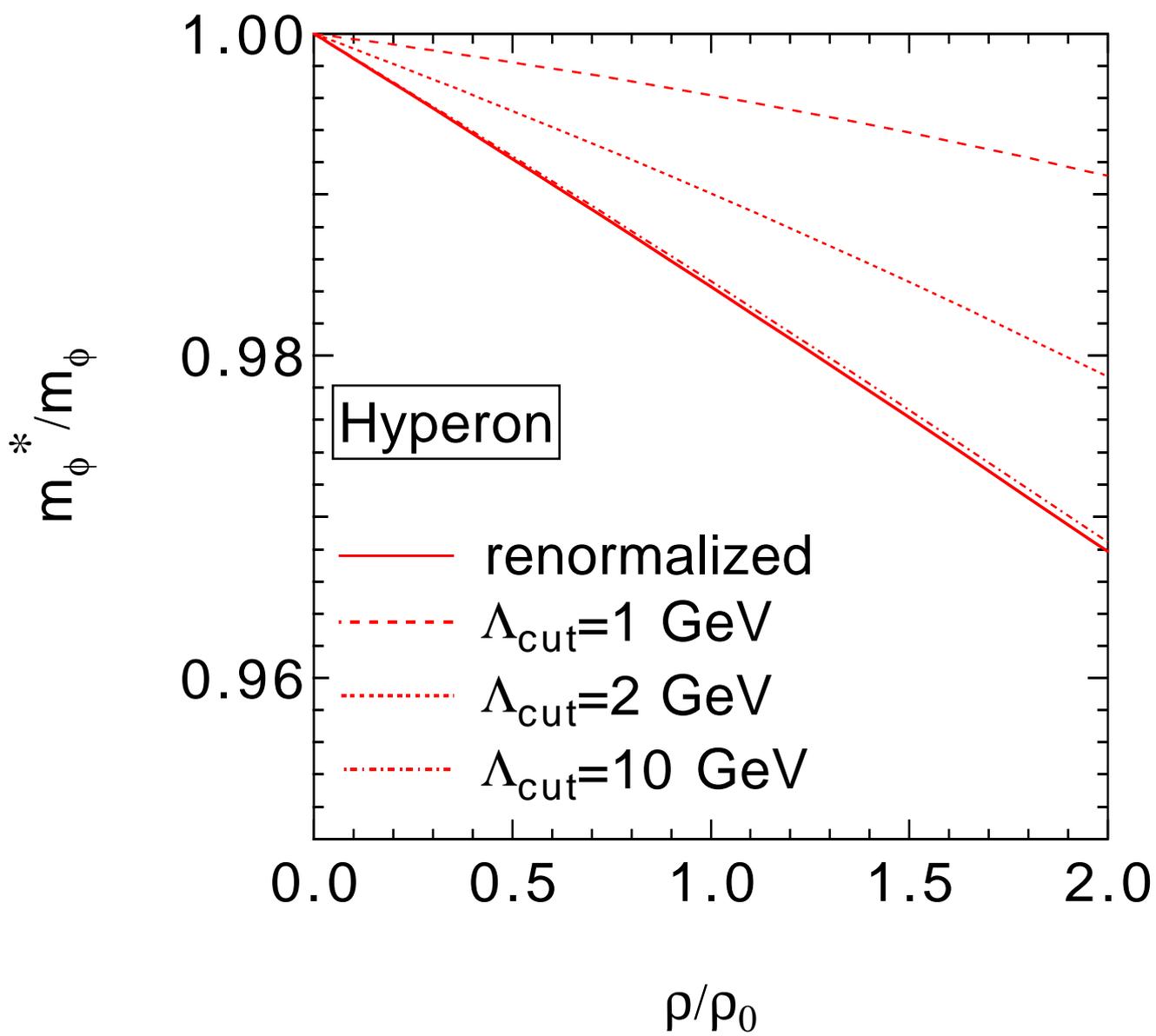
 Fig.2: Ratio of the $\phi$-meson mass in matter $m_{\phi}^*$ and that
 in vacuum $m_{\phi}$ as a function of $\rho/\rho_0$.
 Only hyperon contributions are included in the $\phi$-meson self-energy.

\vspace{0.8cm}

\noindent
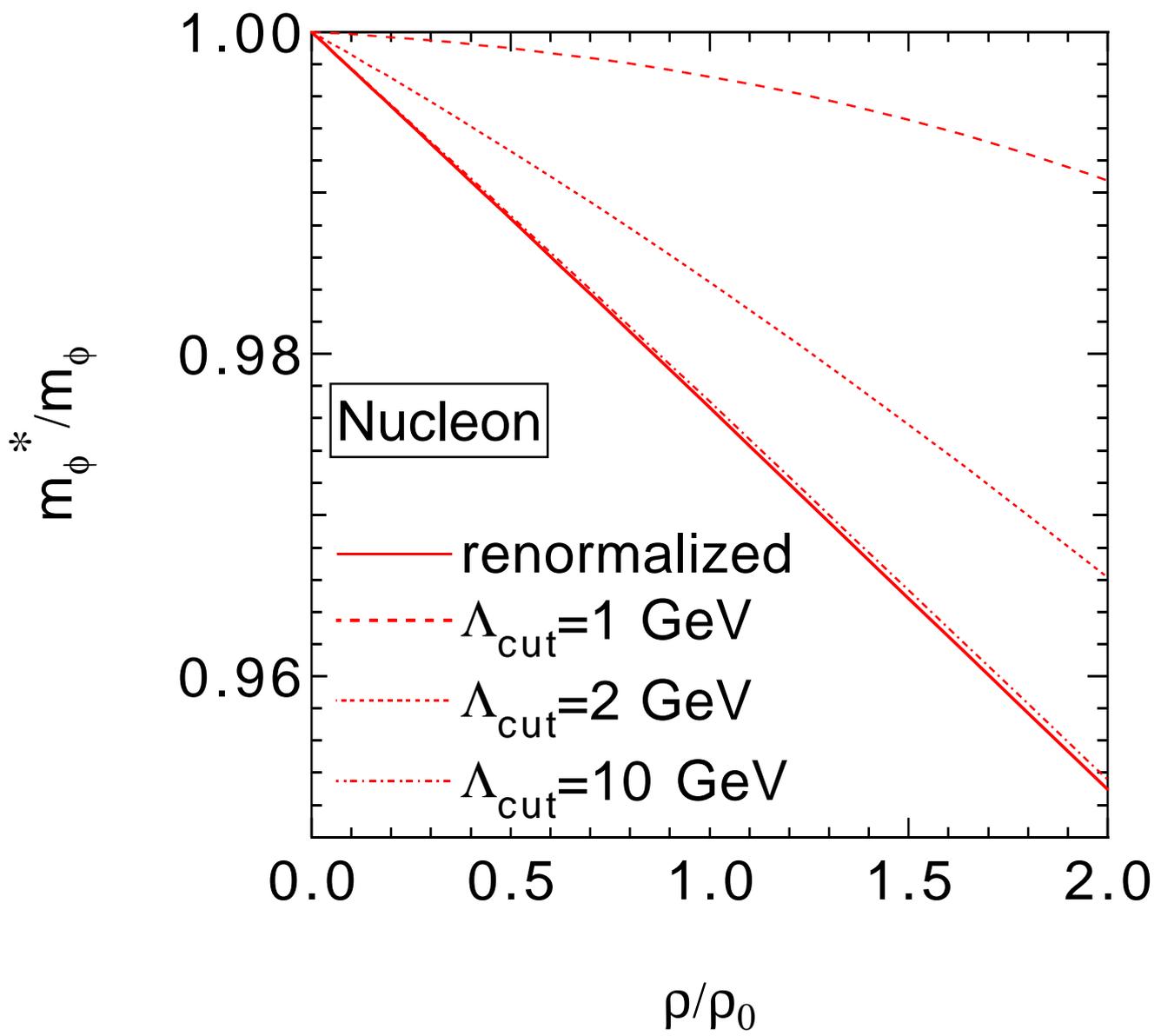
 Fig. 3: Same with Fig.2 except that
 only the nucleon contribution is included.


\newpage


\begin{thebibliography}{99}
\bibliographystyle{unsrt}


\bibitem{yukawa}
 T. Hatsuda, {\em Hadron Structure and the QCD Phase Transition},
  hep-ph/9502345 (1995); R. Pisarski, {\em Applications of Chiral Symmetry},
 hep-ph/9503330 (1995);
  G. E. Brown and M. Rho, {\em Chiral Restoration in Hot and/or
 Dense Matter}, hep-ph/9504250 (1995).

\bibitem{finiteT}
 D. Lissauer and E. V. Shuryak, Phys. Lett. {\bf B253} (1991) 15;\\
P. Bi and J. Rafelski,  Phys. Lett. {\bf B262} (1991) 485;\\
J. P. Blaizot and R. Mendez Galain, Phys. Lett. {\bf B271} (1991) 32.


\bibitem{LH}
T. Hatsuda and S. H. Lee, Phys. Rev. {\bf C46} (1992) R34.




\bibitem{theory}
 C. M. Ko, P. Levai, X. J. Qiu and C. T. Li, Phys. Rev. {\bf C45}
 (1992) 1400.\\
 K. L. Haglin and C. Gale, Nucl. Phys. {\bf B421} (1994) 613.\\
 M. Asakawa and C. M. Ko, Phys. Rev. {\bf C50} (1994) 3064;
  Nucl. Phys. {\bf A572} (1994) 732.



\bibitem{AGS}  B. Cole, in Proceedings of {\em Quark Matter '95},
 Nucl. Phys. {\bf A} (1995) in press.


\bibitem{Enyo}  H. Enyo, in {\em Properties and Interactions
 of Hyperons}, ed. B. Gibson, P. Barnes and K. Nakai
 (World Scientific, 1994, Singapore); KEK-PS proposal (1994).


\bibitem{SH} H. Shiomi and T. Hatsuda, Phys. Lett. {\bf B334} (1994)  281.

\bibitem{others}
 K. Saito, T. Maruyama and K. Soutome, Phys. Rev. {\bf C40} (1989)
 407;\\
 H. Kurasawa and T. Suzuki, Prog. Theor. Phys. {\bf 84} (1990)
 1030;\\
 K. Tanaka, W. Bentz, A. Arima and F. Beck, Nucl. Phys. {\bf A528}
 (1991) 676;\\
 M. Jaminon and G. Ripka, Nucl. Phys. {\bf A564}
 (1993) 505.



\bibitem{french}   J. C. Caillon and J. Labasouque, Phys. Lett.
 {\bf B311} (1993) 19

\bibitem{PW}
 H. -C. Jean, J. Piekarewicz and A. G. Williams, Phys. Rev. {\bf C49}
 (1994) 1981.


\bibitem{QM} See e.g., J. J. J. Kokkedee, {\em The Quark Model},
 (Benjamin, 1969, New York).


\bibitem{gled}  N. K. Glendenning et al., Phys. Rev. {\bf C48}
 (1993) 889.


\bibitem{jaffe}  R. L. Jaffe, Phys. Lett. {\bf B229} (1989) 275.

\bibitem{QHD} B.D. Serot and J.D. Walecka, Adv. Nucl. Phys. {\bf 16} (1986).

\bibitem{KS88} H. Kurasawa and T. Suzuki,
 Nucl. Phys. {\bf A490} (1988) 571.

\bibitem{lipkin} H. Lipkin, Int. J. of Mod. Phys. {\bf E1} (1992) 603.


\bibitem{Yabu} H. Yabu, F. Myhrer and K. Kubodera, Phys. Rev.
 {\bf D50} (1994) 3549, and references therein.





\end{thebibliography}
\end{document}